\documentclass[conference]{IEEEtran}
\IEEEoverridecommandlockouts
\usepackage{cite}
\usepackage{amsmath,amssymb,amsfonts}
\usepackage{algorithmic}
\usepackage{graphicx}
\usepackage{textcomp}
\usepackage{xcolor}
\def\BibTeX{{\rm B\kern-.05em{\sc i\kern-.025em b}\kern-.08em
    T\kern-.1667em\lower.7ex\hbox{E}\kern-.125emX}}
\begin{document}

\title{Monte Carlo neutron transport using low power mobile GPU devices
\thanks{}
}

\author{\IEEEauthorblockN{1\textsuperscript{st} Changyuan Liu}
\IEEEauthorblockA{
\textit{BigCompute Laboratory}\\
Beijing, China \\
changyuan\_liu@163.com, chyliu@umich.edu}
}

\maketitle

\begin{abstract}
The using of GPU for Monte Carlo particle transport is lacking of fair comparisons. This work performs simulations on both CPU and GPU in the same package under the same manufacturing process of low power mobile devices. The experiment with simple pincell benchmark problems with fresh fuel gives consistent results between CPU and GPU. In the meanwhile, it finds that the Apple M1 GPU is as twice capable as M1 CPU, while entitled with a 5 times advantage in power consumption. The particle sorting algorithm optimized for GPU improves computing efficiency by 28\%, while prominently reducing GPU power consumption. Such advantage of sorting algorithm is expected to be greater for depleted fuel problems than fresh fuel problem. The kernel reconstruction Doppler broadening algorithm designed for continuously varying materials is demonstrated to produce consistent Doppler coefficients with the reference code and the algorithm can be efficiently implemented on GPU. Compared with the reference code with double precision floating point numbers, the testing codes with single precision floating point numbers could underestimate the K-effective values by about 500 pcm, and the Doppler coefficients of the fuel are well reproduced though. The conclusion may strengthen the argument that it is helpful for high performance computer to adopt GPU in order to reduce gross power consumption.
\end{abstract}


\section{Introduction}
The Monte Carlo method for neutron transport is an algorithm for high fidelity reactor simulation, but requires large amount of computation. Therefore, the more powerful graphics processing units (GPU) gradually becomes a powerful tool to accelerate the Monte Carlo method. Existing researches \cite{nelson2009}\cite{bergmman2015}\cite{xu2015}\cite{hamilton2019}\cite{choi2020}\cite{tramm2022}\cite{tramm2014a}\cite{tramm2014b}\cite{shriver2019}\cite{ma2019} have optimized the GPU algorithms in many aspects, and achieved significant progresses. One of the important method worthy of mention is the optimization method of particle sorting.

However, it is difficult to answer whether GPU is advantageous over CPU, and by how much the advantage is, because that the manufactures of CPU and GPU may be different or the manufacturing processes are different. Nowadays, extra scale super computers have deployed GPUs in mass scales\cite{frontier2022}, so researches are more and more focusing on whether GPU is advantageous over CPU in terms of computing speed and power consumption.

Therefore, this article adopts the GPU and CPU in the same package and under the same manufacturing process, and simulates small scale neutron transport problems, in order to better compare GPU and CPU. The resultant benefits are the following.
\begin{itemize} 
\item First, low power mobile devices are usually in mass production, while adopting mature chip manufacturing processes, so these devices are suitable for performance measurement related to the specific chip manufacturing processes.
\item Second, low power mobile devices integrate CPU and GPU on the same chip at the same time, where their manufactures are (usually) the same with the same chip manufacturing process, which leads to a fair platform for comparison.
\item Third, low power mobile devices are more easily attainable, which offers convenience to redo the experiment in the artticle.
\item Finally, the experiment with of low power mobile devices offers reference information for whether to set up GPUs in high performance computers (HPCs).
\end{itemize}

\section{Low Power Mobile Devices}
These years, as consumers' demands for mobile computing increase, the computing capabilities of mobile phones, tablets, and skim and thin laptops are improved as well. What worthy of mentioning are system on chips (SoCs) such as Apple M1\cite{apple2020} (Tab. \ref{tab:modbile_devices}), whose computing power approaches desktop computers.

\begin{table}[htbp]
\caption{The summary of mobile low power consumption computing devices}
\begin{center}
\begin{tabular}{|c|c|c|}
\hline
\textbf{Computing devices}&\multicolumn{2}{|c|}{\textbf{Apple M1}} \\
\hline
\textbf{Launch date} & \multicolumn{2}{|c|}{Nov. 2020} \\
\hline
\textbf{Type}& CPU & GPU \\
\hline
\textbf{Endianess}& Little & Unknown\\
\hline
\textbf{Computing units} & 4/4 & 8 \\
\hline
\textbf{Peak frequency (MHz)} & 3,204/2,064 & 1,278 \\
\hline
\textbf{Peak power (W)} & $\sim$15 & $\sim$ 13\\
\hline
\textbf{Memory size (GB)} & \multicolumn{2}{|c|}{16} \\
\hline
\textbf{Memory frequency (MHz)} & \multicolumn{2}{|c|}{4,266} \\
\hline
\textbf{Memory type} & \multicolumn{2}{|c|}{LPDDR4X} \\
\hline
\end{tabular}
\label{tab:modbile_devices}
\end{center}
\end{table}

Low power consumption computing devices usually integrate the CPU and GPU inside the same chip package in order to reduce the size of PCB broad. CPU typically uses the ARM instruction sets designed for low power consumption devices, and the CPU cores are usually unsymmetric in order to balance the need for high performance in short times and long lasting daily usage. Some CPU cores are more powerful and demanding more energy as well, and some have median power while using less energy.

In addition, to meet the requirement of image quality for video games, the GPUs become more powerful, and the GPUs are more favorable in terms of power consumption per unit computation.

\subsection{Apple M1}

\begin{figure}[htbp]
\centerline{\includegraphics[width=\linewidth,trim={5in 0in 5in 0in},clip]{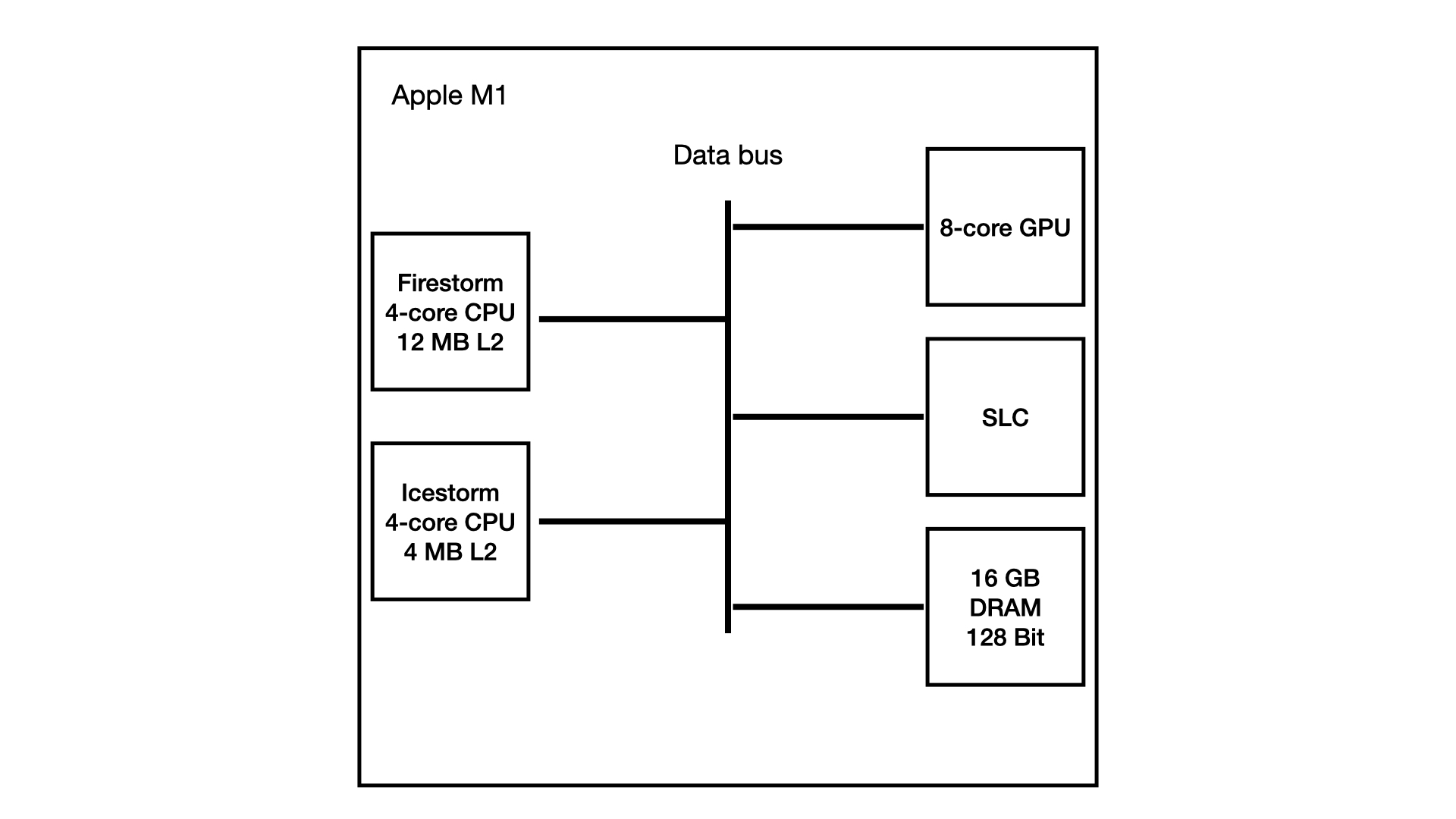}}
\caption{Illustration of Apple M1 package. Firestorm and Icestorm are code names. L2 is the second level cache. SLC is the system level cache. Only parts are included.}
\label{fig:apple_m1}
\end{figure}

Apple M1 SoC integrates the CPU and GPU as indicated by Fig. \ref{fig:apple_m1}. CPU has 8 computing units, among which 4 are more powerful than the other 4. GPU has 8 computing units. The system level cache (SLC) serves all the computing units on chip at the same time.

\section{Monte Carlo neutron transport codes}
In the following, this section introduces the test and reference programs used for neutron transport, whose main characteristics are signified in Tab. \ref{tab:code_features}.

\begin{table}[htbp]
\caption{The main characteristics of test and reference codes}
\begin{center}
\begin{tabular}{|c|c|c|c|}
\hline
\textbf{Codes}& \multicolumn{2}{|c|}{\textbf{Test code}} & \textbf{Reference code}\\
\hline
\textbf{Processor type} & CPU & GPU & CPU\\
\hline
\textbf{Floating point number} &  \multicolumn{2}{|c|}{Single}  & Double \\
\textbf{precision}& \multicolumn{2}{|c|}{} &\\
\hline
\textbf{OTF$^{\mathrm{a}}$ Doppler broadening} &  \multicolumn{2}{|c|}{Kernel reconstruction} & WMP$^{\mathrm{b}}$ turned off\\
& \multicolumn{2}{|c|}{} & (use tabulated)\\
\hline
\textbf{Unresolved resonance} & \multicolumn{2}{|c|}{Turned off} & Turned off\\
\hline
\textbf{Thermal scattering} & \multicolumn{2}{|c|}{Turned off} & Turned off\\
\hline
\textbf{Resonance scattering} & \multicolumn{2}{|c|}{Turned off} & Turned off\\
\hline
\multicolumn{4}{l}{$^{\mathrm{a}}$OTF: on-the-fly.}\\
\multicolumn{4}{l}{$^{\mathrm{b}}$WMP: windowed multipole.}
\end{tabular}
\label{tab:code_features}
\end{center}
\end{table}

\subsection{Test Code}\label{AA}
Since the Monte Carlo neutron transport codes for both mobile CPU and GPU are not available from third party, an in-house experiment code is developed.

In terms of programming design, GPU usually has wider execution pipelines than CPU, but each pipeline has more limited processing power and limited cache. Therefore, device specific optimizations of instructions are required in order to successfully execute programs.

To ensure that the CPU and GPU results are comparable, the input files and nuclear data need to be the same on one side, and the CPU and GPU codes are required to deliver consistent results without the influence of floating point numbers errors. In addition, the CPU or GPU targeted optimization should not harm the consistency of results. Because of the random nature of the Monte Carlo algorithm, the perturbation in the history of a particle may trigger avalanche in the statistics in numerical results. There are researches \cite{tramm2022} to show that the consistency of numerical results between CPU and GPU is attainable under acceptable errors.

Besides, because of the difference in the hardware driver of GPU manufactures and application interfaces, to ensure the numerical consistency across different platforms, the optimization targeting a specific platform deserves careful treatment.

\subsection{Reference Code}
OpenMC\cite{romano2015} is a Monte Carlo particle transport software owned by MIT and Argonne National Laboratory. The choice of OpenMC as a reference code will lead to more persuasive results on one side and meet the need of redoing the results by compiling source code of a specific version on the other side. So, although industrial practice suggests evaluating the results of OpenMC with cautions, OpenMC is suitable for providing reference results. Tab. \ref{tab:openmc_env} lists the building and running environment of OpenMC, in order to redo or evaluate the results. For industrial operations, it suggests to use programs such as JMCT\cite{deng2018}, RMC\cite{wang2015}, MCX\cite{he2021}, cosRMC\cite{qin2018} and etc.

\begin{table}[htbp]
\caption{The building and running environment of the reference OpenMC code}
\begin{center}
\begin{tabular}{|c|c|}
\hline
\textbf{Configuration} & \textbf{Parameters}\\
\hline
\textbf{Code version} & 0.13.1-dev (7752afb, May 16, 2022)\\
\hline
\textbf{Compiler options} & cmake [the location of CMakeLists.txt] \\
& -DCMAKE\_CXX\_STANDARD=17\\
\hline
\textbf{Source of cross section library} & ENDF/B-VIII.0\\
\hline
\textbf{Cross section library processing code} & NJOY 2016.65 (Nov. 1, 2021)\\
\hline
\textbf{Initial random seed} & Default\\
\hline
\textbf{CPU} & AMD 5800X\\
\hline
\textbf{Operating system} & Ubuntu 20.04.2\\
\hline
\end{tabular}
\label{tab:openmc_env}
\end{center}
\end{table}

\subsection{On-the-fly Doppler Broadening}
The high fidelity reactor simulation requires handling with continuously variable temperatures. In earlier researches, the on-the-fly Doppler broadening method named `kernel reconstruction' is demonstrated to be effective\cite{ducru2017}, and suitable for implementing on GPUs, and work well with continuously variable materials\cite{liu2022}. Therefore, the test code adopts the kernel reconstruction method.

As a result, although the temperatures such as 600 K, 900 K and 1,200 K from the benchmark problems are common temperatures prepared by a cross sections data library, the test code does not use them directly. Instead, the test code calculates the Doppler broadened cross sections on-the-fly with kernel reconstruction method. The consequence is that such conclusions are suitable for arbitrary temperatures as well.

Besides, OpenMC provides a collection of test data for the windowed multipole Doppler broadening method\cite{josey2016} based on the ENDF/B-VII.1 library\cite{wmp2022}, but not on the ENDF/B-VIII.0 library.

\subsection{Unresolved Resonance, Thermal Scattering and Resonance Scattering}
In current experiment, the unresolved resonance, thermal scattering and resonance scattering are not taken into considerations. For pressurized water benchmark problems, although it requires innovations in GPU algorithms at times, the percentage of work burden added by the unresolved resonance, thermal scattering and resonance scattering is relative small. \cite{choi2021}

\subsection{Optimization Methods for GPU Monte Carlo}
There are many GPU implementation of Monte Carlo methods\cite{nelson2009}\cite{bergmman2015}\cite{xu2015}\cite{hamilton2019}\cite{choi2020}\cite{tramm2022}\cite{tramm2014a}\cite{tramm2014b}\cite{shriver2019}\cite{ma2019}. Here discusses one important optimization. i.e. the particle sorting. Neutron transport needs to retrieve the cross sections  for all component nuclides under certain particle energies, which accounts for the majority of the computing burdens, but the sorting of particles may relieve this burden. 

\subsubsection{Sorting of Material Types}
A reactor is made of many types of materials such as fuel, burnable poisson, cladding, gap, structural materials and so on, for which there are a large discrepancy in the nuclide compositions. If the neighboring GPU pipelines are processing two different materials, because of the difference in the nuclide being processed, the pipeline can not execute in parallel, but in series instead, so as to prolong the execution time. Hence, the sorting of particles per material types may try its best to ensure the neighboring pipelines to retrieve the same nuclide, and accelerate the execution.

\subsubsection{Sorting of Particle Energies}
In addition, it would degrade the execution speed if two neighboring pipelines have particle energies far from each other.  This is because that the cross section library usually store the data per the order of particle energies, so that the locations of cross sections for near energies are near in memory address as well. When a GPU pipeline retrieves data from memory, it will trigger the cache to keep data in the neighborhood of address requested as well. As a result, the cross sections of nearer energies have higher probabilities to be cached. Therefore, when neighboring pipelines are accessing to cross sections at nearer energies, later pipelines are more probable to find data already in the cache. Moreover, since the access to cache is faster than the access to memory, the program execution is hence accelerated. So sorting per particles' energies will help the program's execution.

\section{Benchmark Problems and Calculation Parameters}
This section describes the benchmark problems and calculation parameters.

\subsection{Benchmark Problems}
The current benchmark problems are a set of problems of simple geometries, including VERA benchmark 1B, 1C and 1D, which are single pin-cell problems with same nuclide compositions but different temperatures in the central fuel regions, i.e. 600 K, 900 K and 1,200 K for the three problems respectively. \cite{godfrey2014} These problems with same material compositions but different temperatures offer a chance to compare Doppler coefficients.

\subsection{Nuclide Data Library}
The benchmark problems have 38 nuclides in total, which takes less than 1GB including the Doppler broadening data at codes' run time. Such size is small enough to fit in the limited memory of low poer consumption mobile devices. Detailed information about the nuclear data library is summarized in Tab. \ref{tab:nuclear_data_library}.

\begin{table}[htbp]
\caption{The characteristics of nuclear data library for benchmark problems}
\begin{center}
\begin{tabular}{|c|c|c|}
\hline
\textbf{Characteristics} & \textbf{Test Code} & \textbf{Reference Code (OpenMC)}\\
\hline
\textbf{ENDF version} & ENDF/B-VIII.0 & ENDF/B-VIII.0\\
\hline
\textbf{Nuclides count} & 38 & 38 \\
\hline
\textbf{Processing code} & In-house & NJOY \\
\hline
\textbf{Runtime data size} & $<$1GB (include OTF DB$^{\mathrm{a}}$) & Unknown (should $<$ 1GB)\\
\hline
\multicolumn{3}{l}{$^{\mathrm{a}}$OTF DB: on-the-fly Doppler broadening.}\\
\end{tabular}
\label{tab:nuclear_data_library}
\end{center}
\end{table}

\subsection{Monte Carlo Transport Calculation Parameters}
Since the CPU computing units of the processors tested are less than 10, relative small number of particles per batch may saturate the computing resources. Here each batch uses about 131 thousands of particles and let the simulation proceed 1,200 batches with first 200 inactive.

For running on GPU, it requires a rather large amount of particles per batch to saturate the computing resources. Here each batch uses about 1.05 millions of particles and let the simulation proceed 800 batches with first 200 inactive.

The reference code uses about 1.05 millions of particles and let the simulation proceed 1,200 batches with first 200 inactive. The detailed running parameters are listed in Tab. \ref{tab:transport_parameters}.

\begin{table}[htbp]
\caption{The parameters of Monte Carlo transport calculation}
\begin{center}
\begin{tabular}{|c|c|c|c|}
\hline
\textbf{Characteristics} & \multicolumn{2}{|c|}{\textbf{Test code}} & \textbf{Reference code}\\
\hline
\textbf{Processor type} & CPU & GPU & CPU \\
\hline
\textbf{Processor} & \multicolumn{2}{|c|}{Apple M1} & AMD 5800X\\
\hline
\textbf{Particles per batch} & 131,072 & 1,048,576 & 1,048,576\\
\hline
\textbf{Total batches} & 1,200 & 800 & 1,200 \\
\hline
\textbf{Inactive batches} & 200 & 200 & 200 \\
\hline
\textbf{Effective particles} & 131,072,000 & 629,145,600 & 1,048,576,000\\
\hline
\end{tabular}
\label{tab:transport_parameters}
\end{center}
\end{table}

\section{Numerical Results}
\subsection{K-effective Values}
K-effective values are calculated using the track-length estimators, where the results are detailed in Tab \ref{tab:keff_results} with one standard deviation.

\begin{table}[htbp]
\caption{The K-effective values of benchmark problems}
\begin{center}
\begin{tabular}{|c|c|c|c|}
\hline
\textbf{Benchmark} & \multicolumn{2}{|c|}{\textbf{Test code}} & \textbf{Reference code} \\
\hline
& M1 CPU (1$\sigma$) & M1 GPU (1$\sigma$) & AMD 5800X (1$\sigma$)\\
\hline
\textbf{1B} & 1.177 24 (13) & 1.176 36 (6) & 1.182 56 (4) \\
\hline
\textbf{1C} & 1.166 27 (13) & 1.166 13 (6) & 1.172 45 (5) \\
\hline
\textbf{1D} & 1.157 66 (12) & 1.157 74 (6) & 1.164 04 (4) \\
\hline
\end{tabular}
\label{tab:keff_results}
\end{center}
\end{table}

Although the difference of K-effective values between the test and reference code is large, the results are consistent across CPUs and GPUs for the test code.

\subsection{Doppler Coefficients}
\begin{figure}[htbp]
\centerline{\includegraphics[width=\linewidth,trim={1in 0in 1in 0in},clip]{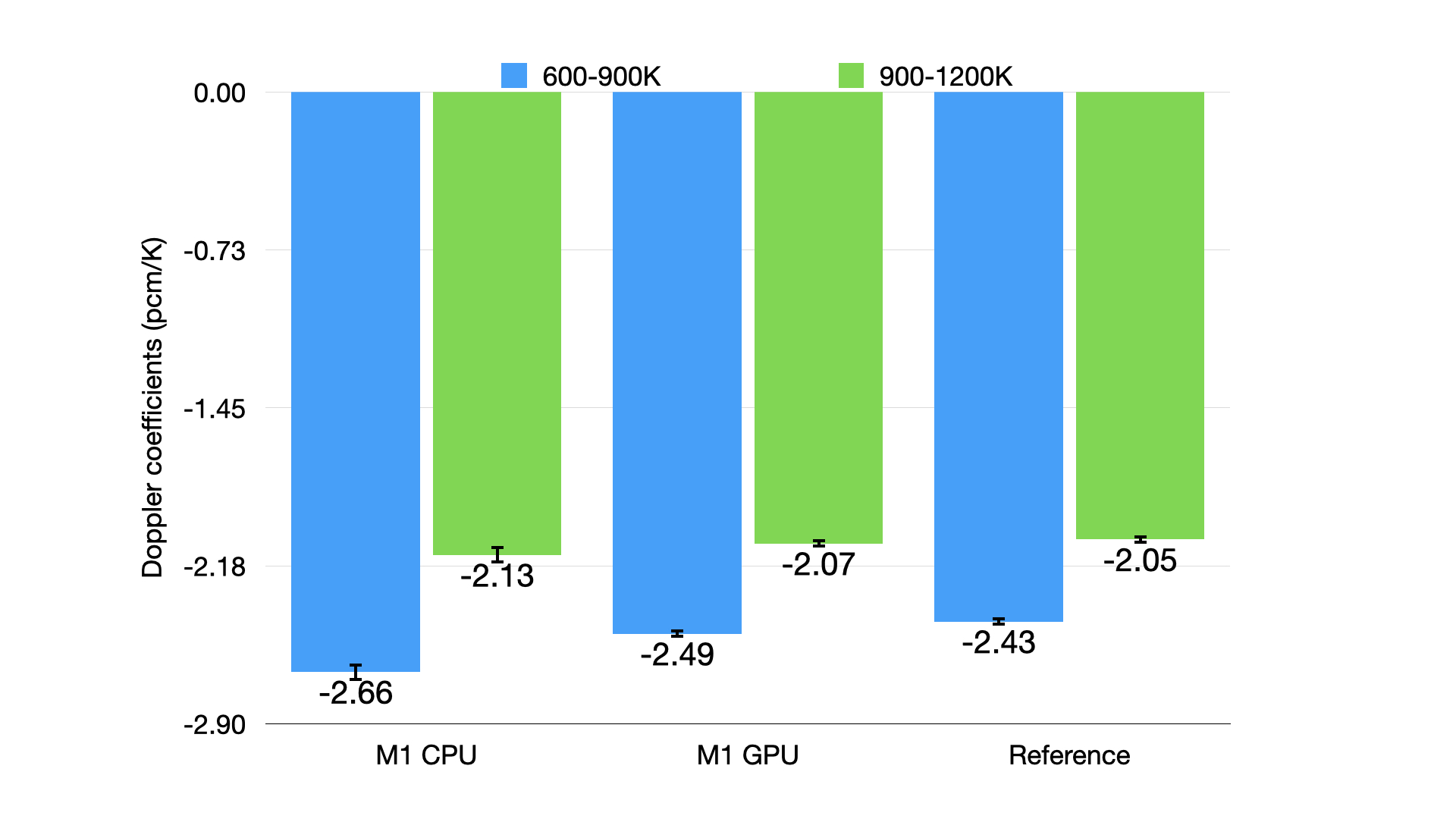}}
\caption{Comparison of Doppler coefficients across different computing devices.}
\label{fig:doppler_coefficient}
\end{figure}

Since the benchmark problems are only different in fuel temperatures and all the nuclide components are the same, the testing of Doppler coefficients may disclose more about the source of bias in K-effective values. Fig. \ref{fig:doppler_coefficient} lists the Doppler coefficients from 600 K to 900 K and from 900 K to 1,200 K, which are consistent between the test and the reference code. The Doppler coefficient from temperature $T_1$ to $T_2$ is defined as \eqref{eq:doppler_coefficient}.

\begin{equation}
\alpha_{T_1\to T_2}=\left(\frac{1}{k_1}-\frac{1}{k_2}\right)\times10^5\mbox{ pcm}/(T_2-T_1),\; T_1<T_2\label{eq:doppler_coefficient}.
\end{equation}

The numerically close doppler coefficients indicate the applicability for the test code in the prediction of key features of nuclear reactors.

Besides, the test code does not use the cross sections at 600 K, 900 K, and 1,200 K, but use the kernel reconstruction method to doppler broaden on-the-fly. This demonstrates again that the kernel reconstruction is effective.

\subsection{Computing Capability and Power Consumption}
\begin{figure}[htbp]
\centerline{\includegraphics[width=\linewidth,trim={1in 0in 1in 0in},clip]{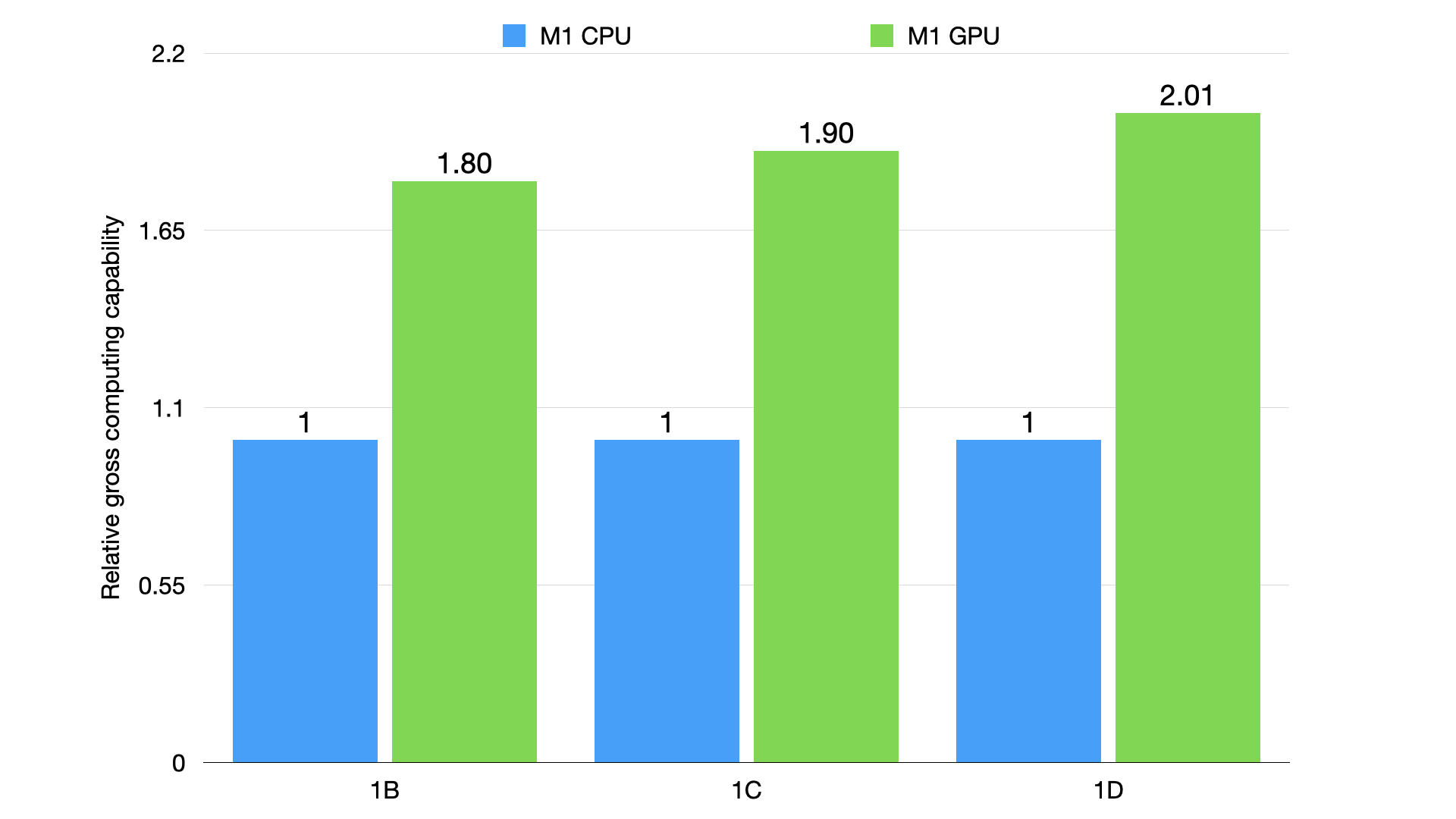}}
\caption{Comparison of total computing power of different computing devices. Use Apple M1 CPU as baseline.}
\label{fig:computing_capability}
\end{figure}

Fig. \ref{fig:computing_capability} compares the relative gross computing capability of different devices, with M1 CPU as the baseline. The computing capability is proportional to the number of particles processed per unit time. The gross computing capability of M1 GPU is about twice of M1 CPU.

\begin{figure}[htbp]
\centerline{\includegraphics[width=\linewidth,trim={1in 0in 1in 0in},clip]{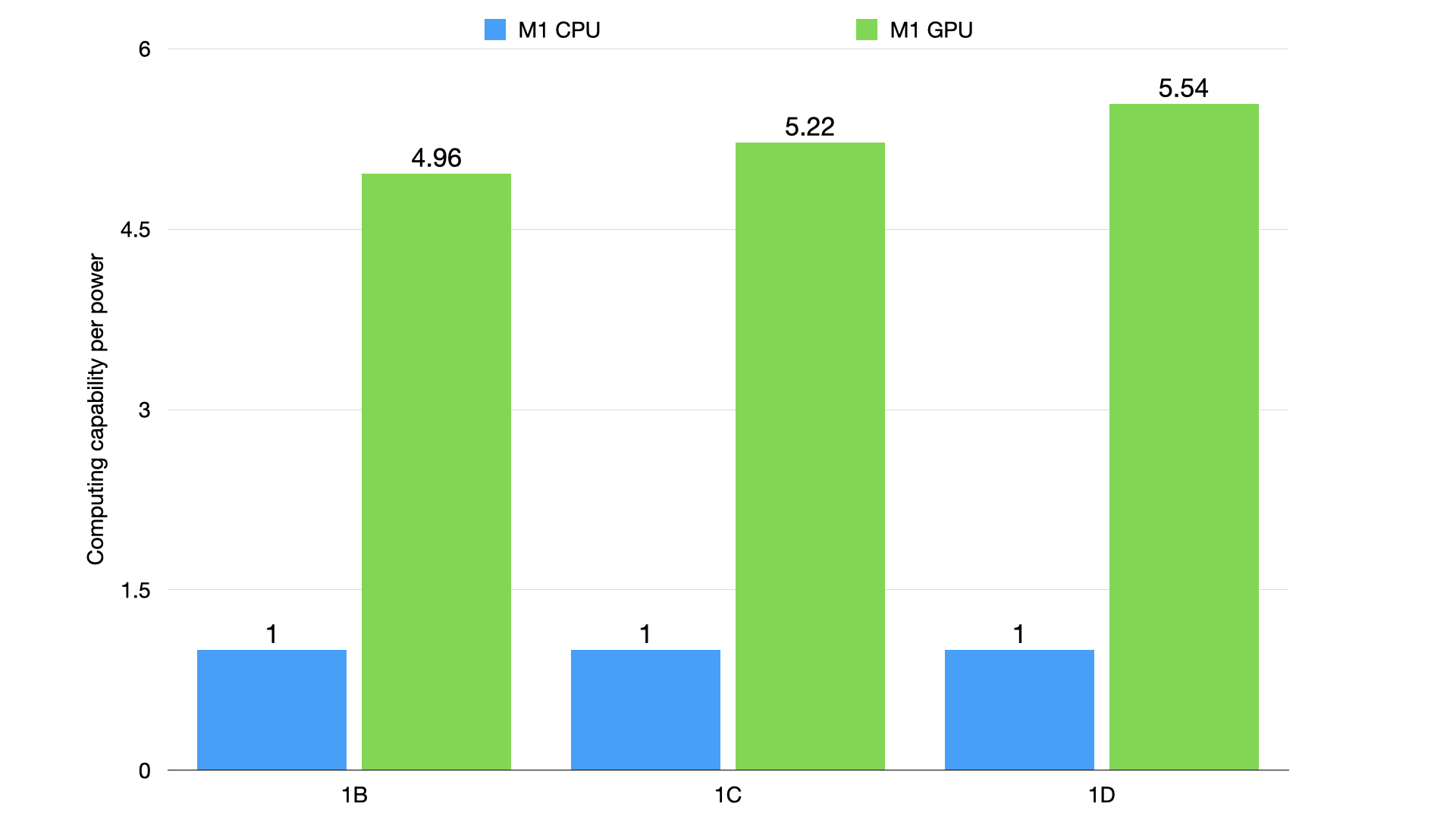}}
\caption{Comparison of total computing power per energy consumption of different computing devices. Use M1 CPU results as a baseline.}
\label{fig:capability_per_power}
\end{figure}

In addition, Tab. \ref{tab:power_consumption} lists the actual power consumption, and Fig. \ref{fig:capability_per_power} compares the computing capability per power consumption of different devices. The M1 GPU leads M1 CPU by a factor of 5 in the computing capability per power.

\begin{table}[htbp]
\caption{Comparison of the power consumption of different computing devices}
\begin{center}
\begin{tabular}{|c|c|}
\hline
\textbf{Computing devices} & \textbf{Power consumption (W)}\\
\hline
\textbf{M1 CPU} & $\sim$11$^{\mathrm{a}}$\\
\hline
\textbf{M1 GPU} & $\sim$4$^{\mathrm{a}}$ (package power)\\
& (including $<$0.5 CPU, $<$0.2 DRAM)\\
\hline
\multicolumn{2}{l}{$^{\mathrm{a}}$Measured with the program \textit{powermetrics}.}
\end{tabular}
\label{tab:power_consumption}
\end{center}
\end{table}

\section{GPU Particle Sorting Strategy}
\begin{figure}[htbp]
\centerline{\includegraphics[width=\linewidth,trim={1in 0in 1in 0in},clip]{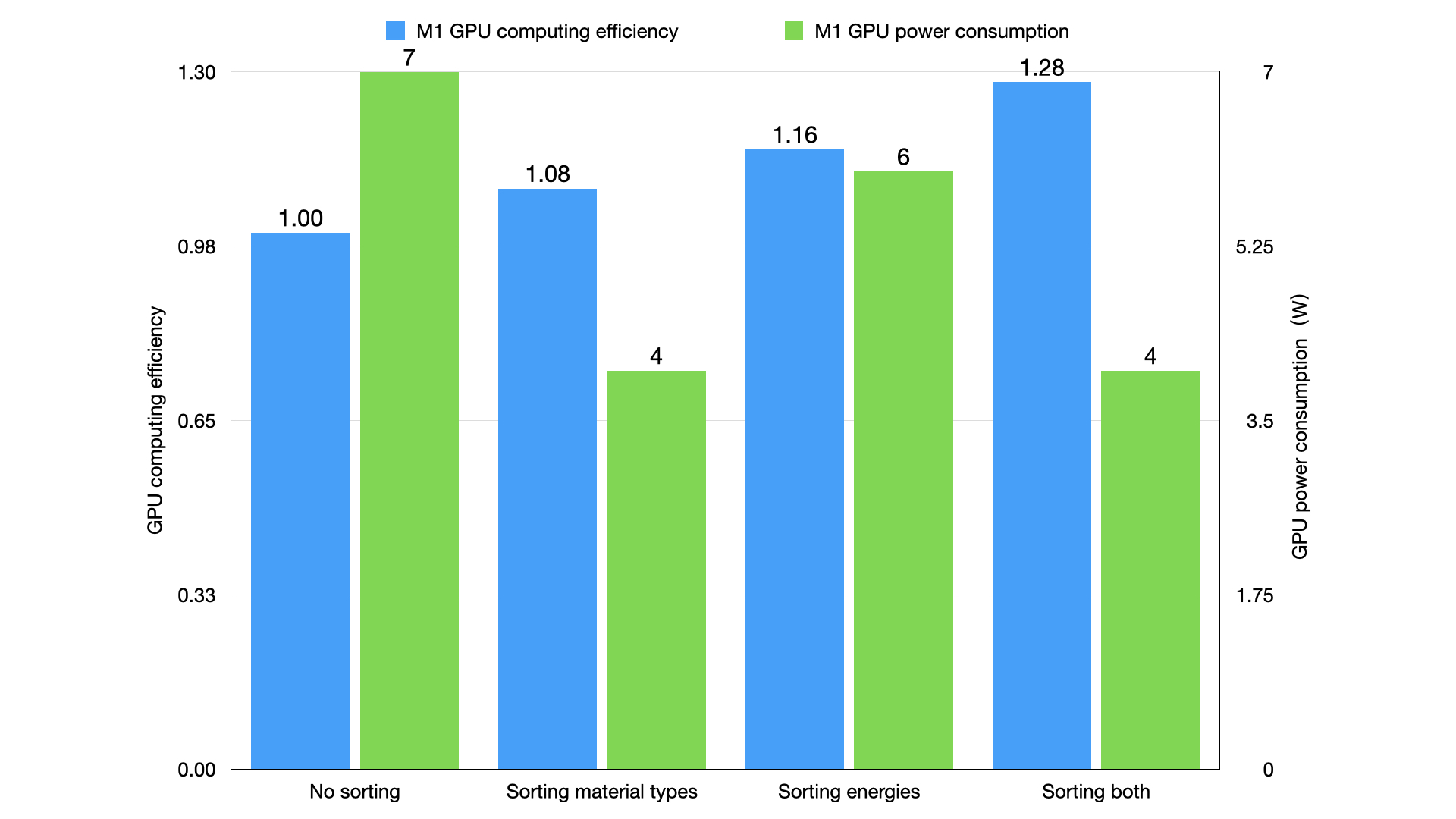}}
\caption{Comparison of the effects of particle sorting strategies on computing efficiency and power consumption (sorting time not included). Use Apple M1 GPU. Use non-sorting results as baseline.}
\label{fig:gpu_sorting}
\end{figure}

Practice has shown that the sorting of particle from the GPU algorithm may promote execution speed and out of one's expectation greatly reduces the power consumption. Fig. \ref{fig:gpu_sorting} presents the computing efficiency and power consumption with different particle sorting strategies, with no sorting results as baseline. The computing efficiency is proportional to the number of particle processed per unit time. The timing excludes the cost of particle sorting, since the reference work does not disclose the sorting algorithm, where some work sorts on CPU, and some work sorts on GPU, which leads to unfair comparison. 

After sorting both the material types and particle energies, the computing efficiencies, proportional to the number of particle simulated per unit time, increases by 28\%. In addition, the sorting of material types reduces the power consumption by 40\%, while the influence of sorting particle energies on power consumption is small. What attributed to this is that the sorting of material types increase the chance that two neighboring pipelines executing at the same time, which hence leads to decrease in the use of execution instruction scheduler to coordinate branches in the execution pipelines. As a result, the gross power consumption is hence reduced by reducing the power consumed by the instruction execution scheduler.

In earlier work by Hamilton et. al.\cite{hamilton2019}, for whole core simulation with fresh fuel, the sorting of materials leads to an increase of execution speed by 1.1 times with the total timing excluding the sorting timing, and for whole core transport with depleted fuel, the promotion in computing speed (excluding sorting timing) is about 1.52 times. The speedup on M1 GPU with sorting materials for fresh fuel pincell is about 1.08, which agrees well with the work by Hamilton et. al..

Moreover, Choi et. al.\cite{choi2021} have observed that the sorting of materials and energy leads to a speedup exclude sorting timing by 1.25 times for a pincell with fresh fuel, and by 4.12 times for a pincell with depleted fuel with 290 nuclides. The speedup on M1 GPU with fresh fuel pincell is 1.28 times, which agrees well with the work by Choi.

In addition, Tramm et. al.\cite{tramm2022} has observed that for whole core simulation problem with depleted fuel, the sorting of materials leads to a speedup of 1.3 on top of the energy sorting. The corresponding speedup on M1 GPU is 1.19 times.

According to these reference work, the advantage of M1 GPU over M1 CPU will be further enlarged for problems with depleted fuel.

\section{Conclusion}

This work first performs Monte Carlo neutron transport on the CPU and GPU inside the same chip package, which provides a fair platform to compare the computing capability and power consumption between CPU and GPU. The computing device under study is Apple M1.

Second, this work simulates the VERA pincell benchmark problems. The test code with single precision floating point numbers underestimates the k-effective values by about 500 pcm compared to the reference code using double precision floating point numbers, where both codes share the same source of cross sections and the same nuclear physics simulation procedures. While as, the test codes on both CPU and GPU lead to consistent results under the same conditions of single precision floating point numbers. 

Third, the test code adopts the kernel reconstruction Doppler broadening algorithm designed for continuously variable materials. Compared to the reference code with the point-wise cross sections from a set of temperatures used in the reference code, the kernel reconstruction method reproduces the Doppler coefficients of fuel.

Fourth, the gross computing capability of M1 GPU is about twice of M1 CPU, and the M1 GPU has an advantage of 5 times per power computing capability over M1 CPU. 

Fifth, the sorting of material types and particle energies may promote computing performance and lower the power consumption. On M1 GPU, the application of both sorting algorithm improves the computing efficiency by 28\%, and the sorting of material types may lower the power consumption by 40\%. The effects of sorting algorithm agrees well with reference work. For problems with depleted fuel, the advantage of GPU may be further enlarged.

However, the results in this work needs more experiments to confirm the conclusions.

In summary, this article leads to the conclusion that GPU may be advantageous over CPU for Monte Carlo particle transport, which may provide a support argument for why super computers will use GPU.

\section*{Acknowledgment}

This work is an English translated version of an article submitted to the CORPHY 2022 conference\cite{corphy2022}, which is postponed to be held in Shanghai in 2023. Original text is in Chinese, and there may be minor difference in the contents.

This work is also an extension to the published article \cite{liu2022}.

Computing technologies from BigCompute Laboratory are used to produce parts of the data in this article. BigCompute Laboratory \& its information providers endeavor to ensure the accuracy \& reliability of the information provided, but do not guarantee completeness or reliability, or that it is up-to-date \& accepts no liability (whether in tort or contract or otherwise) for any loss or damage, whether direct or indirect, arising from errors, inaccuracies or omissions or the information being up-to-date. Any information provided is at the user’s risk.

\end{document}